\documentclass[useAMS, a4, usenatbib]{mn2e}
\usepackage{amsmath,environ}
\usepackage{subfigure}

\input psfig.sty

\usepackage{floatflt, epsfig}
\usepackage{epstopdf}

\def\spose#1{\hbox to 0pt{#1\hss}}
\def\lta{\mathrel{\spose{\lower 3pt\hbox{$\mathchar"218$}}
     \raise 2.0pt\hbox{$\mathchar"13C$}}}
\def\gta{\mathrel{\spose{\lower 3pt\hbox{$\mathchar"218$}}
     \raise 2.0pt\hbox{$\mathchar"13E$}}}
\def\kms{{km s$^{-1}$}}

\def \chisq  {\ifmmode  \chi^2   \else  $\chi^2$  \fi}  
\def \spose#1{\hbox  to 0pt{#1\hss}}  
\def \lta{\mathrel{\spose{\lower 3pt\hbox{$\sim$}}\raise  2.0pt\hbox{$<$}}}
\def \gta{\mathrel{\spose{\lower  3pt\hbox{$\sim$}}\raise 2.0pt\hbox{$>$}}}

\def \kms {\ifmmode  \,\rm km\,s^{-1} \else $\,\rm km\,s^{-1}  $ \fi }
\def \kpc {\ifmmode  {\rm~kpc}  \else ${\rm~kpc}$\fi}  
\def \pc {\ifmmode  {\rm~pc}  \else ${\rm~pc}$ \fi  }  
\def \Gyr {\ifmmode  {\rm~Gyr}  \else ${\rm~Gyr}$\fi}
\def \Msun {\ifmmode \rm{M}_{\odot} \else M$_{\odot}$ \fi} 
\def \Lsun {\ifmmode L_{\odot} \else $L_{\odot}$ \fi} 
\def \Rsun {\ifmmode R_{\odot} \else $R_{\odot}$ \fi} 
\def \Msunpyr {\ifmmode M_{\odot}{\rm~yr}^{-1} \else $M_{\odot}{\rm~yr}^{-1}$ \fi} 
\def \hMsun {\ifmmode h^{-1}\,\rm M_{\odot} \else $h^{-1}\,\rm M_{\odot}$ \fi}

\def \LCDM {\ifmmode \Lambda{\rm CDM} \else $\Lambda{\rm CDM}$ \fi}
\def \sig8 {\ifmmode \sigma_8 \else $\sigma_8$ \fi} 
\def \OmegaM {\ifmmode \Omega_{\rm M} \else $\Omega_{\rm M}$ \fi} 
\def \OmegaL {\ifmmode \Omega_{\rm \Lambda} \else $\Omega_{\rm \Lambda}$\fi} 
\def \Omegab {\ifmmode \Omega_{\rm b} \else $\Omega_{\rm b}$ \fi}
\def \Deltavir {\ifmmode \Delta_{\rm vir} \else $\Delta_{\rm vir}$ \fi}
\def \rhocrit {\ifmmode \rho_{\rm crit} \else $\rho_{\rm crit}$ \fi}
\def \rhou {\ifmmode \rho_{\rm u} \else $\rho_{\rm u}$ \fi}
\def \zc {\ifmmode z_{\rm c} \else $z_{\rm c}$ \fi}

\def \rhos {\ifmmode \rho_{\rm s} \else $\rho_{\rm s}$ \fi} 
\def \rs {\ifmmode r_{\rm s} \else $r_{\rm s}$ \fi} 
\def \cvir {\ifmmode c_{\rm vir} \else $c_{\rm vir}$ \fi} 
\def \Rvir {\ifmmode r_{\rm vir} \else $R_{\rm vir}$ \fi}
\def \Vvir {\ifmmode V_{\rm  vir} \else  $V_{\rm vir}$  \fi} 
\def \Mvir {\ifmmode M_{\rm  vir} \else $M_{\rm  vir}$ \fi}  
\def \Nvir {\ifmmode N_{\rm  vir} \else $N_{\rm  vir}$ \fi}  
\def \Jvir {\ifmmode J_{\rm vir} \else $J_{\rm vir}$ \fi} 
\def \Evir {\ifmmode E_{\rm vir} \else $E_{\rm vir}$ \fi} 
\def \vvir {\ifmmode v_{\rm vir} \else $v_{\rm vir}$ \fi} 
\def \lam {\ifmmode \lambda  \else $\lambda$ \fi} 
\def \lamp {\ifmmode \lambda^{\prime} \else $\lambda^{\prime}$  \fi} 
\def \Vmax {\ifmmode V_{\rm  max} \else  $V_{\rm max}$  \fi} 
\def \Mdm {\ifmmode M_{\rm  dm} \else $M_{\rm  dm}$ \fi}

\def \Mgas {\ifmmode M_{\rm gas} \else $M_{\rm gas}$ \fi} 
\def \Mcg {\ifmmode M_{\rm cg} \else $M_{\rm cg}$\fi} 
\def \Mhg {\ifmmode M_{\rm hg} \else $M_{\rm hg}$ \fi} 
\def \Mdisc {\ifmmode M_{\rm disc} \else $M_{\rm disc}$ \fi} 
\def \Md {\ifmmode M_{\rm d} \else $M_{\rm d}$ \fi} 
\def \Mda {\ifmmode M_{\rm d,0\%} \else $M_{\rm d,0\%}$ \fi} 
\def \Mdb {\ifmmode M_{\rm d,20\%} \else $M_{\rm d,20\%}$ \fi} 
\def \Mdc {\ifmmode M_{\rm d,40\%} \else $M_{\rm d,40\%}$ \fi} 
\def \md {\ifmmode m_{\rm d} \else $m_{\rm d}$ \fi} 
\def \Mb {\ifmmode M_{\rm b} \else $M_{\rm b}$ \fi} 
\def \Mbh {\ifmmode M_{\rm b,pri} \else $M_{\rm b,pri}$ \fi} 
\def \Mbs {\ifmmode M_{\rm b,sat} \else $M_{\rm b,sat}$ \fi} 
\def \zo {\ifmmode z_{0} \else $z_{0}$ \fi} 

\def \rb {\ifmmode r_{\rm b} \else $r_{\rm b}$\fi}
\def \rs {\ifmmode r_{\rm s} \else $r_{\rm s}$\fi}
\def \rc {\ifmmode r_{\rm c} \else $r_{\rm c}$\fi}
\def \rvir {\ifmmode r_{\rm vir} \else $r_{\rm vir}$\fi}
\def \rbh {\ifmmode r_{\rm b,pri} \else $r_{\rm b,pri}$ \fi} 
\def \rbs {\ifmmode r_{\rm b,sat} \else $r_{\rm b,sat}$ \fi}

\title[Local photoionisation feedback]{Galaxy Formation with local photoionisation feedback I. Methods}

\author[Kannan et al.]{R. Kannan$^1$\thanks{Email: kannan@mpia.de}, G.\,S. Stinson$^{1}$, A. V. Macci\`o$^1$, J. F. Hennawi$^1$,  R. Woods$^2$, J. Wadsley$^{2}$,
\newauthor{S. Shen$^3$, T. Robitaille$^1$, S. Cantalupo$^3$, T. R. Quinn$^{4}$, C. Christensen$^5$}
\vspace*{6pt}\\
$^{1}$Max-Planck-Institut f\"ur Astronomie, K\"onigstuhl 17, 69117, Heidelberg, Germany\\
$^{2}$Department of Physics and Astronomy, McMaster University, Hamilton, Ontario, L8S 4M1, Canada\\
$^{3}$Department of Astronomy \& Astrophysics, UCO/Lick Observatory, University of California, 1156 High St, Santa Cruz, CA 95064, USA \\
$^{4}$Astronomy Department, University of Washington, Box 351580, Seattle, WA, 98195-1580, USA\\
$^5$Department of Astronomy, University of Arizona, 933 North Cherry Avenue, Rm. N204, Tucson, AZ 85721-0065, USA}

\begin{document}
\maketitle
\label{firstpage}

\begin{abstract} 

We present a first study of the effect of local photoionising 
radiation on gas cooling in smoothed particle hydrodynamics (SPH) simulations 
of galaxy formation.  We
explore the combined effect of ionising radiation from young and old stellar 
populations. 
The method computes the effect of multiple radiative sources 
using the same tree algorithm used for gravity, so it is computationally
efficient and well resolved.  The method foregoes calculating
absorption and scattering in favour of a constant escape fraction
for young stars to keep the calculation efficient enough
to simulate the entire evolution of a galaxy in a cosmological
context to the present day. This allows us to quantify the effect of the local 
photoionisation feedback through the whole history of a galaxy's formation.
The simulation of a Milky Way-like galaxy using the local photoionisation model
forms $\sim 40$ $\%$ less stars than a simulation that only includes a
standard uniform background UV field. 
The local photoionisation model decreases star formation by
increasing the cooling time of the gas in the halo and
increasing the equilibrium temperature of dense gas in the disc. 
Coupling the local radiation field to gas cooling from the halo 
provides a {\it preventive} feedback mechanism which keeps the central disc 
light and produces slowly rising rotation curves without resorting to 
extreme feedback mechanisms.
These preliminary results indicate that the effect of local 
photoionising sources is significant and should not be ignored in models
of galaxy formation. 
\end{abstract}


\section{Introduction}
Within the current paradigm of galaxy formation theory, 
dark matter first collapses into small haloes, which merge to
form progressively larger haloes.  Galaxies
form out of the gas that cools into the centres of these dark
matter haloes and forms stars  (\citealt{1978MNRAS.183..341W}; \citealt{2010gfe..book.....M}).

Gas collapses into the dark matter haloes by radiating away its 
energy.  Since angular momentum is conserved, the gas settles into a rotating
disc from which stars form.  Radiative 
cooling controls the infall of gas onto the disc, making it one of the most 
important processes of galaxy formation.  
A number of factors slow the infall of gas including thermal pressure 
\citep{1977MNRAS.179..541R,1977ApJ...215..483B} and the incident radiation 
field \citep{1986MNRAS.218P..25R,1992MNRAS.256P..43E,2010MNRAS.403L..16C,2012ApJS..202...13G}.  
Accurately modelling the cooling rate of 
halo gas is critical to determining how much fuel is available to form 
stars in the galaxy.

Gas cooling ($\Lambda$) depends most strongly on gas density 
($\Lambda\sim n^2$), metallicity
and the ionisation state.  The local gas hydrodynamics determines the
gas densities while chemical enrichment from stellar evolution determines
the metallicity.  The gas ionisation state depends on the temperature of the gas 
and on the incident radiation field from stars, Active Galactic Nuclei (AGN) and 
other radiation sources.
In most galaxy formation models (including numerical hydrodynamical simulations), 
a uniform background is used to represent this radiation field.
This background evolves with redshift according to the 
cosmic star formation and quasar luminosity histories \citep{2012ApJ...746..125H}.

\citet{1986MNRAS.218P..25R} and \citet{1992MNRAS.256P..43E} showed that 
photoionisation can prevent gas from cooling into low mass halos.  
\citet{2009MNRAS.393...99W} presented more detailed results that measured 
the effect of a uniform photoionisation background on individual ion species.
\citet{2012ApJS..199...20G} extended this analysis to include a variety
of radiation fields that could vary due to proximity to galaxies. 
\cite{2013MNRAS.434.1063O,2013MNRAS.434.1043O}  explored the problem with a full chemical
network including all the ionisation states of 30 elements in typical parcels
of gas in the intergalactic medium (IGM) and found that the time it takes for
gas to reach ionisation equilibrium can lead to significant changes in the
state of gas in the IGM.

\citet{2010MNRAS.403L..16C} explored analytically the effect of local sources of radiation on the cooling of halo gas including the soft X-ray emission produced by star formation events, a component that is absent within typical stellar population synthesis models such as Starburst99 \citep{1999ApJS..123....3L} (SB99) that only considers the blackbody radiation from young massive stars. Such low energy photons do not affect the cooling rate of 
high metallicity gas. 
However, massive stars are also strong X-ray sources due to their stellar
winds, supernova remnants and binary interactions. 
When the high energy radiation from these sources
is included in gas cooling models, their 
radiation can ionise the metals and can decrease the cooling rate of high 
metallicity gas considerably.
\citet{2012ApJS..202...13G} created a general model for cooling in the presence of a  
radiation field near a galaxy (including both stars and AGNs).  They showed 
that for a sufficiently general variation in the spectral shape and
intensity of the incident radiation field, the cooling and heating functions can 
be approximated based
only the photoionization rates of a few important coolants.

In this paper, we follow the lead of \citet{2010MNRAS.403L..16C} and 
\citet{2012ApJS..202...13G} in an attempt to self consistently 
include local ionisation sources, in addition to the 
uniform background of \citet{2012ApJ...746..125H},
in cosmological simulations of galaxy formation.

One of the great challenges for including the effect of photoionisation in 
simulations is the need to trace the radiation as it propagates through the 
simulated volume.  The radiation field at any given point is dependent on 
the brightness and distance to the source as well as the frequency dependent optical depth of 
the material between the source and sink.  This makes the problem more 
expensive than the $\mathcal{O}(N^2)$ direct calculation of gravity.

Various solutions have been implemented for this complex computational 
problem.  \citet{2008ApJ...673L...1G} used the local Sobolev approximation
that calculates the column density from the density of a resolution element
divided by the size of that element.  \citet{2008MNRAS.386.1931A} (\textsc{sphray}) 
and \citet{2008MNRAS.389..651P} (\textsc{Traphic}) both implemented sophisticated
ray-tracing schemes in smoothed particle hydrodynamics (SPH) simulations.
\cite{2013MNRAS.434..748A} presented a recent update to \textsc{SPHray}.
\citet{2011MNRAS.412..935P} traced radiation through AREPO, a code that
solves hydrodynamics on a moving mesh. 
For a review of how the different schemes perform in a variety of
common test cases, see \citet{2009MNRAS.400.1283I}.  The codes all show that
reionisation of the Universe happens in a non-uniform manner. While such radiative transfer schemes are useful tools for studying
reionisation, these methods are so computationally demanding that it is 
impossible to evolve a cosmological simulation of galaxy formation
much past $z=4$. 

Such models have been used in galaxies simulated to $z=0$ in post-process.  \citet{2011MNRAS.418.1796F} solved radiative transfer on a high resolution grid to find that local radiation ionises low column gas, but has little effect on the statistics of Lyman limit and Damped Lyman alpha systems. \citet{2013MNRAS.431.2261R} used {\sc traphic} post-process and found similar results. However, these studies do not yet explore the impact of the radiation field on the galaxy evolution. 

Since it is as yet unclear what the effect of including local ionisation sources  on galaxy evolution, we have decided to take a simple approach to the calculation
of the radiative transfer, where possible.  Our aim is to find a compromise
between simulating a galaxy in a cosmological context from high redshift down to $z=0$
and the precision of an on-the-fly radiative transfer calculation.

In a companion paper, Woods et al. (in prep) will present the details
of the radiative transfer methods, which is here summarized in section \ref{sec:cs}.
This paper describes how we calculate the cooling rates using that radiative transfer 
method and presents a preliminary simulation based on them.  
The paper is organized as follows:  \S \ref{sec:cooling}
presents the details of the cooling calculation.  
\S \ref{sec:ionsources} describes the photoionisation
sources we explicitly consider in our calculation. 
 \S \ref{sec:UVfield} outlines the approximations used in our radiative 
transfer approach while  \S \ref{sec:coolingtable} describes the construction of the 
cooling table. Finally, in section  \S \ref{sec:test} and  \S \ref{sec:simulation} we present
the results of our implementation of the local photoionisation feedback on a test gas 
particle and on a fully cosmological simulation of galaxy formation.
Our conclusions are presented in  \S \ref{sec:conclusion}.

\section{Gas Cooling}
\label{sec:cooling}
A number of processes determine the internal heating (H) and cooling ($\Lambda$) rates in the hydrodynamic energy equation:

\begin{equation}
\frac{D\epsilon}{Dt} = -\frac{P}{\rho} \vec{\nabla}.\vec{u} - \frac{1}{\rho} \vec{\nabla}.\vec{F} + \frac{1}{\rho}\Psi + \frac{\rm{H}-\Lambda}{\rho}
\label{eq:heat}
\end{equation}
where $\epsilon$ represents the specific internal energy of a parcel of gas,
$P\vec{u}$ represents the adiabatic work done on that gas,
$\vec{F}$ represents the flux of heat that is conducted out of the parcel, 
and $\Psi$ represents the viscous dissipation rate.  

Both the heating and 
cooling rates are a function of the density, $n_i$, of each ion species present
in the gas parcel,
as well as the parcel's temperature, $T$, and incident radiation 
field, $J_\nu$:

\begin{equation}
\frac{H-\Lambda}{\rho} = f(n_i, T, J_\nu)
\end{equation}

The density of each ion species is subject to a 
number of creation and destruction processes, which in turn also depend on
$n_j$, $T$ and $J_\nu$: 
\begin{equation}
n_i = f(n_j, T, J_\nu), j\ne i.
\end{equation}

The densities of the ion species can be obtained using networks of differential
equations that account for all the electrons that are made available when atoms
are ionised.  Such networks can become arbitrarily complicated depending
upon how many elements are included (\citealt{1998PASP..110..761F}; \citealt{ 2012ApJS..199...20G}; \citealt{2013MNRAS.434.1063O}).  
In simulations, it is possible to carry the ionisation state
of every species from timestep to timestep to keep the non-equilibrium 
ionisation state of each element rather than making the assumption of 
 ionisation equilibrium (eg. \citealt{2013MNRAS.434.1043O}).   
 
\subsection{Primordial Cooling: non-equilibrium}
\label{sec:primordial}
Since these differential equations need to be solved for each particle during
every timestep in a simulation, it becomes necessary to limit the calculation
of non-equilibrium ionisation states to hydrogen and helium.
We use the implementation of \citealt{2010MNRAS.407.1581S}
(see also \citealt{2013arXiv1305.2913V}).  Hydrogen
and helium are the most abundant elements in the Universe, so their temperature
and ionisation state are important in determining the dynamics and cooling of
gas in simulations. 

Primordial gas contains various ionisation states of hydrogen and helium: 
$i \in ({\rm HI, HII, HeI, HeII, HeIII,e^-})$. The rate of change of densities of these species are obtained by solving the following set of differential equations:
\begin{equation}
\begin{array}{lll}
\frac{dn_{\rm{HI}}}{dt} = \alpha _{\rm{HII}} n_{\rm{HII}} n_{e} - \Gamma _{e\rm{HI}} n_{e}n_{\rm{HI}} - \Gamma _{\gamma \rm{HI}} n_{\rm{HI}} \\ \\
\end{array}
\label{eq:HI}
\end{equation}
\begin{equation}
\begin{array}{lcl}
\frac{dn_{\rm{HeI}}}{dt} &= &(\alpha _{\rm{HeII}} + \alpha_d) n_{\rm{HeII}} n_{e} \\ & & - \Gamma _{e\rm{HeI}} n_{e}n_{\rm{HeI}} - \Gamma _{\gamma \rm{HeI}} n_{\rm{HeI}} \\ \\
\end{array}
\label{eq:HeI}
\end{equation}
\begin{equation}
\begin{array}{lcl}
\frac{dn_{\rm{HeII}}}{dt} &= &\alpha _{\rm{HeIII}} n_{\rm{HeIII}} n_{e} + \Gamma _{e\rm{HeI}} n_{e}n_{\rm{HeI}} \\ & &+ \Gamma _{\gamma \rm{HeI}} n_{\rm{HeI}}  - (\alpha _{\rm{HeII}} + \alpha_d) n_{\rm{HeII}} n_{e} \\ & &- \Gamma _{e\rm{HeII}} n_{e}n_{\rm{HeII}}  - \Gamma _{\gamma \rm{HeII}} n_{\rm{HeII}} \\ \\
\end{array}
\label{eq:HeII}
\end{equation}
where $\alpha_{i} $ is the \emph{radiative} recombination coefficient for 
ion species $i$, $\alpha_{d}$ is the \emph{dielectric} recombination 
coefficient, which only applies to HeII,
$\Gamma_{ei}$ is the \emph{collisional} ionisation rate for each species,
while $\Gamma_{\gamma i}$ is the photo-ionisation rate, defined as
\begin{equation}
\Gamma_{\gamma i} = \int_{\nu_{Ti}}^\infty \frac{4 \pi J_{\nu}}{h\nu} \sigma_{\nu i} \, d\nu
\label{eq:gamma}
\end{equation}
where $\sigma_{\nu i}$ is the frequency dependant  photoionisation cross section of the species `i'.

These equations are closed when combined with the following set of conservation 
equations:
\begin{equation}
n_{\rm{HI}} + n_{\rm{HII}} = n_{\rm{H}}
\end{equation}
\begin{equation}
n_{\rm{HII}} + n_{\rm{HeII}} + 2n_{\rm{HeIII}} = n_{e}
\end{equation} 
The cosmic production of helium was constrained by \citet{2003Sci...299.1552J} 
using K dwarfs from Hipparcos catalog with spectroscopic metallicities and 
found that the amount of Helium produced compared to heavier elements in 
stars follows the relation: $\Delta Y/ \Delta Z = 2.1 \pm 0.4$. In accordance 
with this result we compute the helium abundance in the following manner:  
\begin{equation}
Y_{\rm{He}} = 
\begin{cases}
 (0.236 + 2.1\rm{Z})/4.0 & \text{if } \rm{Z} \le 0.1 \\
 (-0.446(Z - 0.1)/0.9 + 0.446)/4.0 & \text{if } \rm{Z} > 0.1
 \end{cases},
\end{equation}
with the density of hydrogen being
 \begin{equation}
Y_{\rm{H}} = 1.0 - 4Y_{\rm{He}} - \rm{Z},
\end{equation}
where 
\begin{equation}
Y_i = \frac{n_iM_{\rm{H}}}{\rho}
\end{equation}
where $M_H$ is the mass of the hydrogen atom and $\rho$ is the total density of the gas.
Most of the radiative processes discussed above along with collisional 
excitation that causes the gas to cool, and the coefficients and the 
cooling rates are enumerated in \citet{1997NewA....2..209A}. 

In addition to determining the ionisation state of gas, the 
incident radiation field also injects energy into the gas when a high energy 
($h\nu > h\nu_{T}$) photon, transfers the rest of its energy to the electron it frees from
the atom. The photo heating rate is thus given by
\begin{equation}
{\rm H} = n_{\rm{HI}}\epsilon_{\rm{HI}} + n_{\rm{HeI}}\epsilon_{\rm{HeI}} + n_{\rm{HeII}}\epsilon_{\rm{HeII}},
\end{equation}
where
\begin{equation}
\epsilon_i = \int_{\nu_T}^\infty \frac{4 \pi J_{\nu}}{h\nu} \sigma_{\nu i}(h\nu-h\nu_T) \, d\nu
\label{eq:epsilon}
\end{equation}

\subsection{Metal cooling: equilibrium}
\label{sec:mce}
A significant amount of cooling also occurs via elements heavier than hydrogen
and helium.  For these elements, we refrain from solving the entire non-equilibrium
ionisation network due to its computational complexity.  Instead, we assume equilibrium 
conditions hold and interpolate values in a
look-up table as implemented in \citet{2010MNRAS.407.1581S}.  The table consists of heating and 
cooling rates as a function of total gas density, temperature, redshift,
and the radiation fields from local sources, described in \S 
\ref{sec:ionsources}.  Metallicity does not need to be a dimension in our table, since the 
cooling rate scales linearly with metallicity and can thus be easily calculated from the
cooling rate of solar metallicity gas  (see also \citealt{2010MNRAS.407.1581S} 
and \citealt{2013arXiv1305.2913V}).

Thus, our total cooling calculation is the summation of three components:  i) 
non-equlibrium primoridal - calculated on the fly in the code; ii) equilibrium metals - values tabulated from  {\sc Cloudy} (v10.00, last described in \citealt{1998PASP..110..761F}) ; and iii) Compton scattering of CMB photons,
\begin{equation}
\Lambda_{tot} = \Lambda_{\rm{H,He}} + \frac{\rm{Z}}{\rm{Z}_{\odot}}\Lambda_{\rm{Z}_{\odot}} + \Lambda_{c}.
\label{eq:cooltot}
\end{equation}

This division neglects some sources of free electrons in the cooling 
calculation.  {\sc Cloudy} calculates the cooling rates of metals using the 
free electrons created in ionisation equilibrium primordial cooling.
This number of free electrons may be different from the non-equilibrium 
primoridal cooling in our code. 
Additionally, our non-equilibrium primordial cooling calculation assumes 
that the number of free electrons released from metals is negligible.  
Thus, our cooling calculation is a first approximation that can be improved by tracking
non-equilibrium metal cooling in addition to H $\&$ He. 
It is worth noting that this assumption is used
in practically all current numerical implementations of metal dependent gas cooling (\citealt{2010MNRAS.407.1581S}; \citealt{2013arXiv1305.2913V}).

\section{Ionising stellar radiation sources}
\label{sec:ionsources}
There are many radiation sources which produce high energy photons 
and ionise gas in the galaxy. In the following sections we outline the sources 
that we consider in our photoionisation model.
One source of radiation not included in our model is
quasars, since we do not follow the formation
or gas accretion onto super massive black holes in our cosmological
simulations. However, \citet{2013arXiv1305.2913V} showed that the 
low frequency duty cycle of radiation from AGNs result in a 
minimal impact on the large scale gas dynamics in galaxies.

\begin{figure}
\begin{center}
\includegraphics[scale=0.47]{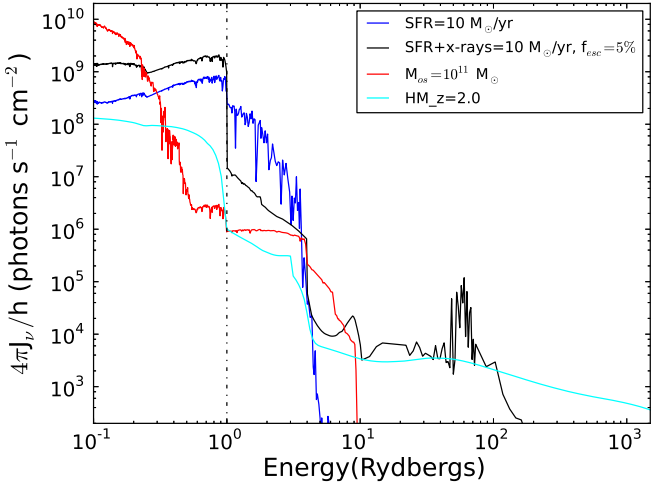}
\caption{The incident photon flux from the photoionisation sources
considered in our simulations.  These fluxes are reported  at
$10$ \kpc \ from the source. The blue curve shows the photon flux as a
function of energy from a population of young stars 
forming at the rate of 10 M$_\odot$ yr$^{-1}$.  The young stars have 
the highest photon flux at the hydrogen edge but decreases quickly at
higher energy.  The  black curve shows the spectrum of young stars 
including the x-ray luminous cooling of shock-heated gas after a SNe event. 
The Spectral Energy Distribution (SED) of a $10^{11}$ M$_\odot$, 
old ($>$200 Myr) stellar population is shown in red.  
Old stars emit fewer photons than young stars at the hydrogen edge, 
but has higher flux at higher energies ($h\nu > 4$ Rydberg).  
The cyan curve shows the UV background at $z=2$.
Compared to the plotted ionisation fields, this background flux is lower 
than local sources at low energies( $h\nu <  100 $Rydberg),
but dominates at higher energies due to the hard photons emitted by AGN.}
\label{fig:sedcomp}
\end{center}
\end{figure} 

\subsection{UV Background}
Nearly all simulations include the effect of photoionisation from a uniform
background.  This UV background accounts for the UV radiation that all stars 
and AGN emit throughout the evolution of the Universe attenuated by the 
Lyman-$\alpha$ forest \citep{2012ApJ...746..125H}.  Our refined method is an attempt
to account for more local ionising radiation.
The Spectral Energy Distribution (SED) from \citet{2012ApJ...746..125H} 
(henceforth, HM) is shown as the cyan curve in Fig. 
\ref{fig:sedcomp}.  Since the HM SED incorporates the emission from AGNs, it 
contains photons up to x-ray energies, but the photon flux can be lower than
local sources at energies less than $10$ Rydbergs.  We consider the HM SED as the minimum
ionising flux seen by the gas particles in our simulations, at low redshift ($z<9$).

\subsection{Young Stars}
\label{ssec:ys}
Our initial spectral energy distribution (SED) for young stars comes 
from \textsc{starburst99} \citep{1999ApJS..123....3L} using an SED 
taken 10 Myr after stars start forming at a constant rate of 
1 M$_\odot$ yr$^{-1}$ using the ``present day'' (Eq. 6) IMF from
\citet{2001MNRAS.322..231K} (blue curve in Fig. \ref{fig:sedcomp}).  The SED has a relatively high photon flux at the 
hydrogen edge ($13.6$ eV) but the flux drops precipitously to higher
energies, with almost no flux above the helium edge (4 Rydbergs $= 54.4$ eV).

\begin{figure}
\begin{center}
\includegraphics[scale=0.47]{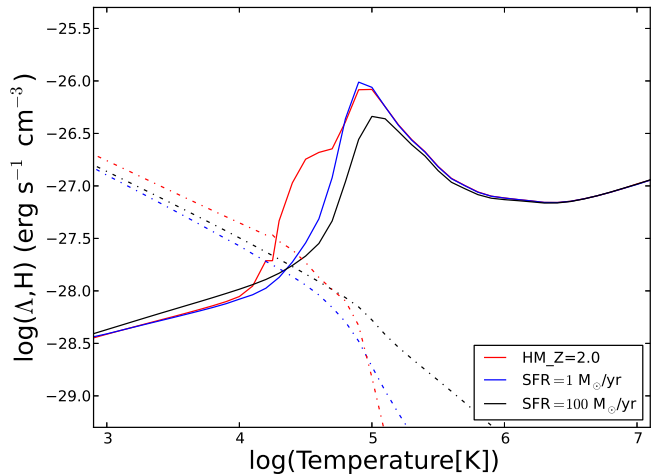}
\caption{The cooling (solid) and heating (dashed) rate curves of gas in
the blackbody radiation field of young stars (\textsc{starburst99}) at 
two star formation rates compared to the cooling and heating in 
presence of the HM UV background spectra.
The cooling rates are shown for gas with a density, $n_H = 0.01$ cm$^{-3}$,
metallicity, $Z = 0.01 Z_{\odot}$, and a radiation source $10$ kpc distant.}
\label{fig:sfrudep}
\end{center}
\end{figure}

Fig. \ref{fig:sfrudep}, shows how the cooling (solid curves) and heating (dashed curves) rates change in the 
presence of various levels of the radiation field (SFR=1 M$_\odot$ yr$^{-1}$, 
blue curve ; SFR=100 M$_\odot$ yr$^{-1}$,  black curve) from young stars compared to 
the cooling and heating rates in the presence of the peak HM radiation 
field (red) at $z=2$.  The gas used to make these
cooling curves has a density of $n_H = 0.01 cm^{-3}$, a metallicity of 
$Z=0.01 Z_\odot$ and is at a distance 10 kpc from the star forming region. 
The cooling curve is calculated using the assumption
that the gas in between the source and the gas test particle is optically thin. 
Thus, the flux is inversely proportional to the of the distance squared. 
All test cooling curves presented in this section were calculated using the code {\sc Cloudy},
which assumes photoionisation equilibrium for all elements.

The \textsc{starburst99} young star SED suppresses hydrogen cooling that 
dominates in $\sim 10^4$ K gas.  The spectrum also partially ionises 
some helium in strong radiation fields. It also slightly changes the equilibrium temperature of the gas.
The equilibrium temperature is defined as the temperature where cooling 
transitions to heating because of the incident radiation file.
Practically,  it  corresponds to the temperature at which the heating 
and cooling rates are equal. 
At temperatures above equilibrium, the gas cools, 
while at lower temperatures, it heats up.  So, the equilibrium temperature 
depends on the shape of the heating and cooling curves.  

Radiation fields
change the heating and cooling curve depending on the energy of photons they
possess.  HM has the most high energy photons, whose extra energy results in
heating, so it has the highest heating rates, but is has a low flux at low energies ($1-10$ Rydbergs) causing minimal impact on the
cooling of low metaliicity gas. The young star blackbody spectrum includes mostly photons
around the hydrogen ionisation edge, so it results in minimal heating, but
a large reduction in the cooling rate. The net result is that the radiation from new stars, for typical values of radiation field in the galaxy, is more effective at raising the equilibrium temperature of the gas than the background HM UV spectrum.

\begin{figure}
\begin{center}
\includegraphics[scale=0.47]{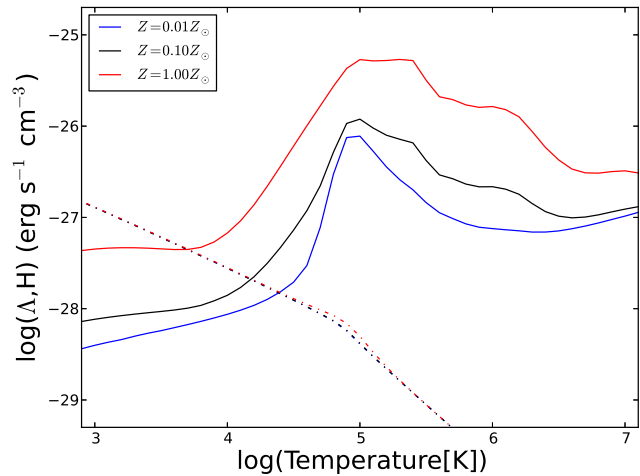}
\caption{Cooling function as in Fig. \ref{fig:sfrudep} of gas at a 
variety of metallicities in a radiation field 10 kpc away from a stellar 
population forming stars at 10 M$_\odot$ yr$^{-1}$.  This young population
only emits radiation to the helium edge (4 Rydberg), so only the hydrogen 
cooling at $T<10^5$ K is reduced.  The increase in cooling in higher 
metallicity gas in the range $10^4<T<10^5$ is due to an increasing presence of heavy metal coolants such as O, Ne and Fe. The heating rate remains constant because the radiation field is the same for all three curves.}
\label{fig:sfrzdep}
\end{center}
\end{figure}

Fig. \ref{fig:sfrzdep} shows the effect of metallicity on the cooling function 
at a fixed star formation rate of 10 M$_\odot$ yr$^{-1}$.
While this radiation field eliminates hydrogen cooling at solar 
metallicity (note the lack of a peak around $10^4$ K in the red curve), 
heavier elements make cooling rates more than an order of magnitude 
higher than in the low metallicity gas (blue) at all temperatures. 
The higher cooling rates mean that the equilibrium temperature of the gas 
is also an order of magnitude lower in solar metallicity (red curve) gas 
because the heating rate does not change.
Fig. \ref{fig:sfrzdep} shows that cooling is most easily 
suppressed in low metallicity gas irradiated with a soft UV spectrum 
from young stars.

\subsection{X-rays from Young Stars}
\label{sec:xrayssource}
\citet{2010MNRAS.403L..16C} considered the radiation from young stellar
populations, including both the blackbody radiation from hot young stars
and the X-rays that supernovae remnants emit.  Using analytic calculations, 
\citet{2010MNRAS.403L..16C} showed that while stellar photons ionise 
hydrogen, soft X-rays ionise other significant metal coolants. 
The x-ray photoionisation increased the equilibrium temperature, which 
consequently slowed the accretion of gas onto the disc.  

X-rays are produced in a number of ways.  Rapidly outflowing gas 
from stellar winds or supernova explosions shocks against the 
interstellar medium (ISM) and thermalises (\citealt{1995ApJ...448...98H}; 
\citealt{2004ApJS..151..193S}). Non-thermal processes associated with 
supernovae explosions and high mass X-ray binaries also emit x-ray radiation 
(\citealt{2003MNRAS.339..793G}; \citealt{2004A&A...419..849P}).

We use the SED for a 5 Myr old stellar population 
(Fig. \ref{fig:sedcomp},  black curve) from \citet{2002A&A...392...19C}.
The \citet{2002A&A...392...19C} SEDs are derived from models of 
young O and B stars and the X-rays that their stellar winds and supernova
explosions produce. Their models
are calibrated to match the observed relationship between SFR and 
soft X-rays (for eg. see \citealt{1995ApJ...448...98H}). In their models, the X-ray emission peaks when the stellar
population is $\sim5$ Myr old and continues for 
$\sim100$ Myr. To simplify our calculation, we use the SED of a 5 
Myr old stellar population for all stars younger than 10 Myr, so we pick the
maximum emission SED, but only use it for one-tenth of the time that
x-rays are emitted.  Following \citet{2010MNRAS.403L..16C}, we assume that 
5\% of the mechanical energy from the SNe is emitted as X-rays.

The \citet{2002A&A...392...19C} models assume a Salpeter IMF, but our 
simulations use a \citet{2003PASP..115..763C} IMF that has
more stars with $M_\star>8$ M$_\odot$.  So, we renormalise the 
\citet{2002A&A...392...19C} SED to make the flux from O $\&$ B stars and 
the number of subsequent SNe events consistent with the Chabrier IMF. 

There is significant absorption of Lyman continuum photons in the galaxy 
due to the abundance of hydrogen. To include this effect we assume an 
escape fraction of 5\% around Lyman-limit frequencies to mimic 
the highly absorptive nature of their birth molecular clouds 
(e.g. \citet{2006A&A...448..513B}, see section \ref{ssec:EF}
for a more thorough discussion on the escape fraction).  

\begin{figure}
\begin{center}
\includegraphics[scale=0.47]{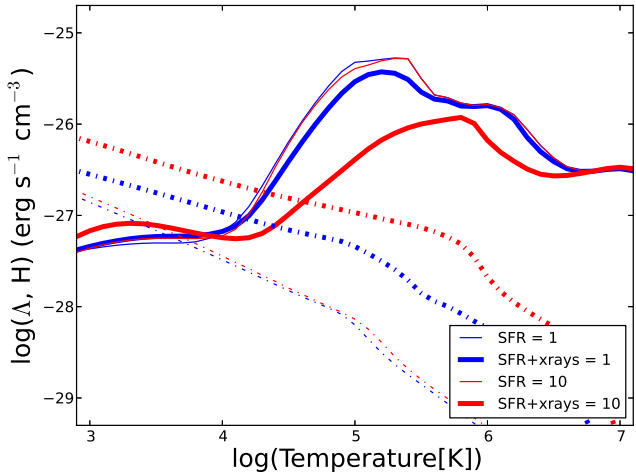}
\caption{The cooling function as in Fig. \ref{fig:sfrudep}, but now includes 
the effect of x-rays from young stellar populations (\citealt{2002A&A...392...19C}). While blackbody 
radiation from young stars only quenches hydrogen cooling, the addition of
x-rays stops cooling from many metal ion species and increase the heating rate 
considerably (see \citealt{2010MNRAS.403L..16C}). }
\label{fig:sfrxudep}
\end{center}
\end{figure} 

Fig. \ref{fig:sfrxudep} shows how including X-rays in the young star SED 
affects the cooling curve in $Z=Z_\odot$ gas at $10^{-2}$ $cm^{-3}$ 
density at $1$ kpc from the star formation site. 
At high metallicity, the young star SED without X-rays (thin curves) only 
eliminates hydrogen cooling, independent of the star formation rate of the galaxy. The harder 
x-ray spectrum (thick curves) ionises heavier elements such
as oxygen, neon and some ions of iron to lower the gas cooling rate 
and also raise the heating rate of the gas. These effects combine to increase
the equilibrium temperature by an order of magnitude.
X-rays thus remain an important ionisation source at low redshifts, 
where the halo gas has been metal enriched by continuous bursts of 
star formation. 
For this reason, we use an SED that combines the black body emission of 
young stars with the x-ray emission that massive stars produce in stellar winds 
and supernova explosions in our simulations. Henceforth, flux from young stars
includes x-rays along with their black body spectrum. 

\subsection{Old Stars}
\label{sec:oldstars}

Naively, old stellar populations seem like they should not be sources of ionising photons.
Since all the hot young stars have exploded as supernovae, all that are left
are cool, old stars.  However, a UV upturn was observed coming from the old
stellar population at the center of M31 \citep{1969PASP...81..475C}, and was determined to
be light from extreme horizontal branch stars.  Subsequently, UV
radiation has been detected from many quiescent, early type galaxies 
\citep{2007ApJS..173..619K}.  What fraction of this radiation is from young stars,
forming at low rates, compared to how much is from old stars is still a question
waiting to be answered with better observations.  Additionally, recent UV telescopes like GALEX
have only been able to detect relatively soft UV radiation sources.  
However stellar population synthesis models predict that stars that have shed their outermost envelopes,
so called ``post-AGB'' stars, should emit a hard UV spectrum.

\citet{2003MNRAS.344.1000B} include such stars in the spectral energy 
distribution of a simple stellar population (SSP).
The SED of SSPs older than $200$ Myr is harder, though much fainter (at the hydrogen edge),
than the UV SED for a young SSP.  In the model, the photon 
flux increases near the helium edge, due to the accumulation of 
post AGB stars.  The shape of the SED remains fairly constant from $200$ Myr 
to $13$ Gyr because low mass stars evolve within a narrow temperature range 
all the way from the main sequence to the AGB phase. This implies that a 
constant SED can be used for old stars irrespective of their age 
(Fig. \ref{fig:sedcomp}, red curve).  We choose the 2 Gyr SED since it 
is near the mean of the SEDs. 

\begin{figure}
\begin{center}
\includegraphics[scale=0.47]{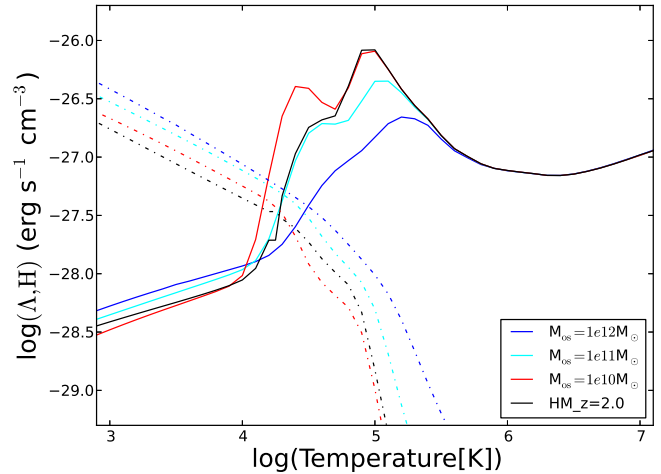}
\caption{ The heating and cooling functions for gas as in Fig. 
\ref{fig:sfrudep} irradiated by various masses of old (t$>$200 Myr) stars. The SED is taken from \citet{2003MNRAS.344.1000B}.}
\label{fig:osudep}
\end{center}
\end{figure}

Fig. \ref{fig:osudep} shows the effect of the old star SED on the cooling 
function of gas as a function of intensity of the incident radiation field. 
The gas has the same conditions as that studied in \S \ref{ssec:ys} 
($n_H = 0.01$ cm$^{-3}$, $Z = 0.01 Z_{\odot}$, d=$10$ kpc). 
The radiation with energies higher than the helium edge can eliminate helium
cooling in the gas. A galaxy with M$_\star=10^{10}$ M$_\odot$ has less effect on 
the cooling function than the peak HM UV background ($z=2$) because its 
ionising flux is so low.  $10^{11}$ \Msun stars produce 
a strong enough radiation field to reduce cooling below the HM level. 
Thus, radiation from old stars starts to play a role in massive galaxies,
where the radiation field is strong, and at lower redshifts, once the HM
background has decreased from its peak.  Thus, they might help limit star 
formation in massive elliptical galaxies.

The consequences of photoionisation that are most critical for galaxy formation
are the decrease in the cooling rate of the gas and the increase in the equilibrium 
temperature, which is the minimum 
temperature to which gas can cool.  Moreover, 
the equilibrium temperature sets the minimum pressure 
a gas parcell can reach.  More intense and harder radiation fields shift the equilibrium 
temperature higher, while higher densities and metallicities shift the 
equilibrium temperature lower.

\section{Calculating the Radiation Field}
\label{sec:UVfield}

In the previous section, the different sources of local ionising radiation 
were enumerated and shown to be important in the calculation of gas cooling 
in galaxies. However, propagating the radiation from the sources to the gas 
particles can become computationally expensive 
(e.g. \citealt{2008MNRAS.386.1931A}).  This necessitates the use of
some simplifying, albeit physically motivated, assumptions. 
We plan to relax these assumptions in future work in order
to present a more complete implementation of radiative transfer.

\subsection {Escape Fractions}
\label{ssec:EF}

Our strongest assumption is that the gas is optically thin to photons 
so that the optical depth is determined simply by the escape fraction of 
ionising photons from the interstellar medium (ISM).  Given the abundance of
neutral hydrogen in the Universe, Lyman continuum photons with around 1 Rydberg
of energy will generally see high optical depths and thus have short travel 
distances and low escape fractions.  In future work, we 
will try to improve our model to make more physical calculations for the 
optical depth.  For now, we assume that since young stars form embedded 
in molecular clouds, the photoionising escape fraction, $f_{esc}$, is 
low ($f_{esc}\sim5\%$). This number is an upper limit, motivated by a number of observations (\citealt{2006A&A...448..513B}; \citealt{2009ApJS..181..272G}; \citealt{2001ApJ...546..665S}; \citealt{2006ApJ...651..688S}; \citealt{2007ApJ...668...62S}; \citealt{2013ApJ...765...47N}). 
The escape fraction we use is frequency dependent ($f^\nu_{esc}$ ), 
similar to \citet{2010MNRAS.403L..16C},
\begin{equation}
f^\nu_{esc} = [f^{LL}_{esc} + (1 - f^{LL}_{esc}) e^{-\tau_\nu}]
\end{equation} 
where $f^{LL}_{esc}$ is the absolute
escape fraction at the Lyman Limit, $\tau_\nu = \sigma_\nu N (H^0 )$ is 
the neutral hydrogen optical depth and $\sigma_\nu$ the corresponding 
cross-section.
Our 5\% escape fraction means that we fix $f^{LL}_{esc}$ = 0.05. The value
of N(H$^0$ ) determines the hardening of the spectrum around the
Lyman Limit. We anticipate that this parameter has a little effect
on our results and we fix N(H$^0$) = 10$^{20}$ cm$^{−2}$ . Escape fractions are 
a strong function of the distance from the source.
$f_{esc}=5\%$ represents the mean escape fraction at our spatial resolution 
$\sim300$ pc. For old stars, we 
assume that they have left their dense birth locations and that the escape 
fraction of their ionising photons is 100\%. 

\subsection{Combining Sources}
\label{sec:cs}

Each parcel of gas will receive a ionising flux from {\it all}
the radiation sources (stars in our case) with an intensity
proportional to the inverse of the distance square.
In order to reduce the computational cost of the distance calculation, 
we exploit the tree algorithm used to compute the gravitational
force 
, which already groups sources according 
to the their distance from the gas particle.

Our grouping scheme is a first rough attempt at calculating radiative
transfer.  Woods et al. (in prep) will present the radiation transfer method
in more detail.
Our initial attempt takes little account of absorption except for the 
constant escape fraction used for young stars described above. 

Combining sources of the same kind is an algebraic operation, since the SEDs do not evolve with time, so the summation of SEDs
only affects their normalisation.  In this manner, the radiation field incident on a gas particle 
is calculated using
two separate components, one from young stars ($\phi_{\rm{SFR}}$) and one from old stars ($\phi_{\rm{os}}$)
Young stars in our scheme are all those with an age $<10$ Myr, while old stars have 
an age $>200$ Myr.\footnote{No flux from stars in between these two ages is considered, in agreement with 
results from \citet{2003MNRAS.344.1000B}}
For each gas particle, the effective flux, $\phi$, is the sum of the sources normalized
by the distance from the gas particle squared, as follows:

\begin{equation}
\phi_{\rm{SFR}} = \frac{1}{10^7 \ \rm{yr}}\sum_{i=1}^N \frac{M_i(t < 10 \rm{Myr})}{({r_i} \ \rm{kpc})^2}
\label{eq:phisfr}
\end{equation}

\begin{equation}
\phi_{\rm{os}} = \sum_{i=1}^N \frac{M_i(t > 200 \rm{Myr})}{({r_i \ \kpc})^2}
\label{eq:phios}
\end{equation}
with the total photoionisation and heating rates, for each species `i', given by

\begin{equation}
\Gamma_{\gamma i} = \phi_{\rm{SFR}}\Gamma_{\gamma i,\rm{SFR}} + \phi_{\rm{os}}\Gamma_{\gamma i,\rm{os}} + \Gamma_{\gamma i,\rm{HM}}
\label{eq:gammatot}
\end{equation}

\begin{equation}
\epsilon_i = \phi_{\rm{SFR}}\epsilon_{i,\rm{SFR}} + \phi_{\rm{os}}\epsilon_{i,\rm{os}} + \epsilon_{i,\rm{HM}}.
\label{eq:epsilontot}
\end{equation}
Here, $\Gamma_{\gamma i,\rm{SFR}}$ and $\epsilon_{i, \rm{SFR}}$ are normalised to a radiation 
field $1$ kpc away from a population forming stars at a rate of 
1 M$_{\odot}$ yr$^{-1}$. These quantities are calculated using  Eqs. \ref{eq:gamma} 
and \ref{eq:epsilon}, with the SED of these stars taken from 
\citet{2002A&A...392...19C} (see \S \ref{sec:xrayssource}). 
Similarly, $\Gamma_{\gamma i,\rm{os}}$ and $\epsilon_{i, \rm{os}}$ are normalised based
on a radiation field $1$ kpc away from a $>200$ Myr stellar population with
a mass of $1$ $M_{\odot}$. For the old stars the SED is taken from 
\citet{2003MNRAS.344.1000B} as mentioned in \S \ref{sec:oldstars}. 
$\Gamma_{\gamma i,\rm{HM}}$ and $\epsilon_{i,\rm{HM}}$ 
are the UV background photionising and photo heating rates, which are 
redshift dependent and taken from \citet{2012ApJ...746..125H}. The cooling table uses 
separate values for $\phi_{\rm{SFR}}$, $\phi_{\rm{os}}$, and redshift in addition to temperature and density to determine the
cooling rate for the gas.

Therefore, the photoionisation rate,
$\Gamma$, and the heating rate, $\epsilon$, for each species ($i$), used in 
the non-equilibrium calculation can simply be summed as shown 
in Eq. \ref{eq:gammatot} $\&$ \ref{eq:epsilontot}. 
The details about how the distance ($r_i$) is calculated will be presented in 
Woods et al. (in prep) (For a short description see \S \ref{sec:simulation}).

\section{Cooling table creation}  
\label{sec:coolingtable}

As shown in Eq. \ref{eq:cooltot}, the total gas cooling is divided into 
primordial, metal and Compton cooling. The primordial and Compton cooling is calculated 
on-the-fly as described in \S \ref{sec:primordial}. 
However, to reduce the complexity of the cooling calculation as the simulation
runs, the metals are assumed to be in ionisation equilibrium and their heating and cooling rates are tabulated 
across a range of physical conditions using \textsc{Cloudy}.  This look 
up table is used in the simulations.  

The table has four dimensions at $z>9$ and five dimensions thereafter.
The four common dimensions are density, temperature, 
$\phi_{\rm{SFR}}$ and $\phi_{\rm{os}}$.  After $z=9$, the UV background turns
on so a redshift dimension is added to track how
the HM SED shape changes.  
We make the division at $z=9$, even though \citet{2012ApJ...746..125H} tabulate 
the UV background to $z=15$, in order to limit
the size of the cooling table.  The mean free path of photons at 
high redshift is low because the Universe is not yet reionised, so the flux 
of the background UV field is relatively low.  There is never a metallicity
dimension in the table as cooling is assumed to scale linearly with metallicity as 
discussed in \S \ref{sec:mce}. 

In both parts of the cooling table, the density ranges from 
$10^{-9}$ $cm^{-3}$ to $10^{4}$ $cm^{-3}$ with a spacing of 0.5 dex in 
log space. The temperature ranges from $10^2$ K to $10^9$ K with a 
resolution of 0.1 dex. 
At $z>9$, $\phi_{\rm{SFR}}$ ranges 
from $10^{-11}$ to $10^{2}$ M$_\odot$ yr$^{-1}$ kpc$^{-2}$ with a resolution 
of 0.5 dex, and $\phi_{\rm{os}}$ ranges from $10$ to $10^{10}$ M$_\odot$ kpc$^{-2}$ 
also with a resolution of 0.5 dex. These minimum values correspond 
to the CIE cooling rates, while 
the maximum values correspond to the maximum star formation rate and mass of
high redshift galaxies. 
For $z\le9$, $\phi_{\rm{SFR}}$ covers a smaller range, 
$10^{-5}$ to $10^{3}$ M$_\odot$ yr$^{-1}$ kpc$^{-2}$ with 
a resolution of 0.5 dex.  $\phi_{\rm{os}}$ ranges from $10^{6}$ to $10^{12}$ M$_\odot$ 
kpc$^{-2}$ with 0.5 dex spacings.  The minimum values are 
higher at $z\le9$ because the HM UV background sets the 
minimum rather than collisional ionisation equilibrium.  
The maximum SFR and mass are increased to reflect observations. 
The redshift dimension that accounts for the UV background ranges from 
9.0 to 0.0 with a resolution of 0.5. 
The resolutions were motivated by \citet{2012ApJS..202...13G}.

To create the table, cooling and heating rates
are calculated at every point for solar and primordial metallicity gas using 
\textsc{Cloudy}.  The difference 
between the solar and primordial metallicity
values is stored as the heating and cooling rates due to the metals only.
The values are stored as natural logarithms
with single point precision, which provides accurate values, and limits
the size of the table to $78$ MB, a size well within the memory 
capacity of modern computer hardware. 

\section{Test Particle Evolution}
\label{sec:test}

The methods described in the previous sections are implemented
in the smoothed particle hydrodynamics (SPH) code {\sc gasoline}
first described in \citet{2004NewA....9..137W}. As a first test of the cooling implementation,
we calculate the evolution 
of the temperature and ionisation state of a single isolated particle over the
course of 350 Myr.  The particle stays at a constant density and metallicity 
since it is not part of a fully dynamic simulation.  The effects of dynamics
are explored in a full simulation in \S \ref{sec:simulation}. 

\begin{figure}
\begin{center}
\includegraphics[scale=0.4]{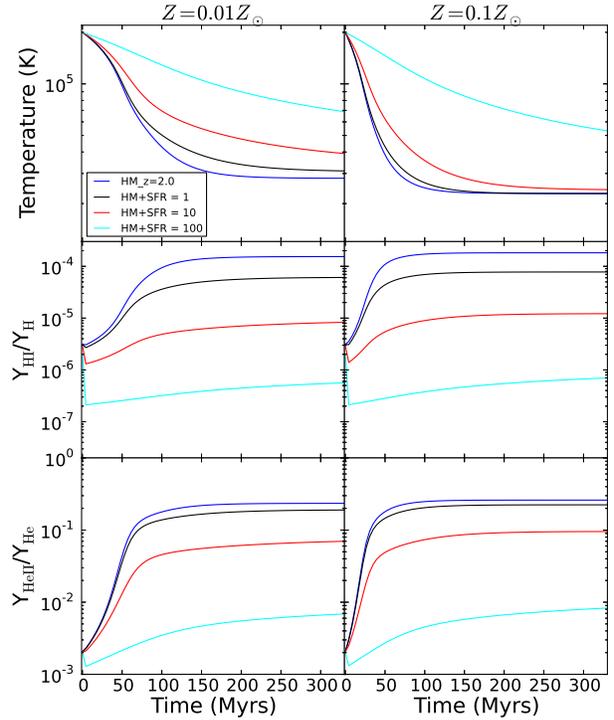}
\caption{ The effect of various star formation radiation fields on 
a parcel of gas with $n_H = 0.001$ cm$^{-3}$ density. The gas cooling 
time and equilibrium temperatures increase significantly even 
for a high metallicity gas due to the presence of high energy x-ray photons.}
\label{fig:sfrx}
\end{center}
\end{figure}

Fig. \ref{fig:sfrx} shows the evolution of a gas particle with a density
$n_H = 0.001$ cm$^{-3}$
in a young star radiation field that includes x-rays 
as described in \S \ref{sec:xrayssource}.  Unlike for the cooling curves
in \S \ref{sec:ionsources} that showed the effect of the stellar fields and 
HM separately, now their fields are combined as described in \S \ref{sec:cs}.  
The effect of the background UV field alone is shown as the blue curve.
The distance from the source to the test gas particle is $10$ \kpc.  The left 
column shows how the particle
cools when its metallicity is 0.01 $Z_\odot$, similar to unenriched gas falling
for the first time into a galactic halo \citep{2013arXiv1306.5766B}.  The 
right column shows the particle cooling time when its metallicity is 0.1 $Z_\odot$, similar 
to gas that has cooled into the disc and has been ejected into the halo
\citep{2013arXiv1306.5766B}.

A comparison of the top two panels shows that cooling times are longer 
and the equilibrium temperatures are higher in low metallicity gas.  
Compared to the HM line, the top right panel shows that
X-rays also prolong cooling at $0.1 Z_\odot$, though to a lesser extent
than at $0.01 Z_\odot$.  
The lower four panels show that the young star radiation fields keep hydrogen and 
helium ionised even after the gas cools.  
What is not shown in the ionisation plots is that they can 
also ionise other potential metal coolants.

 For each metallicity, stronger radiation fields reduce the neutral hydrogen and HeII
fractions (middle and bottom panels).  In the $0.01$ $\rm{Z}_{\odot}$ case, the lack 
of metals means that the delayed cooling is primarily due to the ionisation of 
hydrogen and helium.  While the fractions of neutral hydrogen and HeII 
are similar between the $0.01$ $\rm{Z}_{\odot}$
and $0.1$ $\rm{Z}_{\odot}$ with the same ionisation field, the 
cooling rate in $0.1$ $\rm{Z}_{\odot}$ is higher.  The faster cooling is due to
metal coolants more prevalent in the $0.1$ $\rm{Z}_{\odot}$ gas.  
Since metal cooling is computed using equilibrium gas in \textsc{Cloudy}, it is
impossible to show the ionisation state of the metal coolants in our 
current test particle runs \citep[but see][]{2013MNRAS.434.1063O}. 

\begin{figure}
\begin{center}
\includegraphics[scale=0.4]{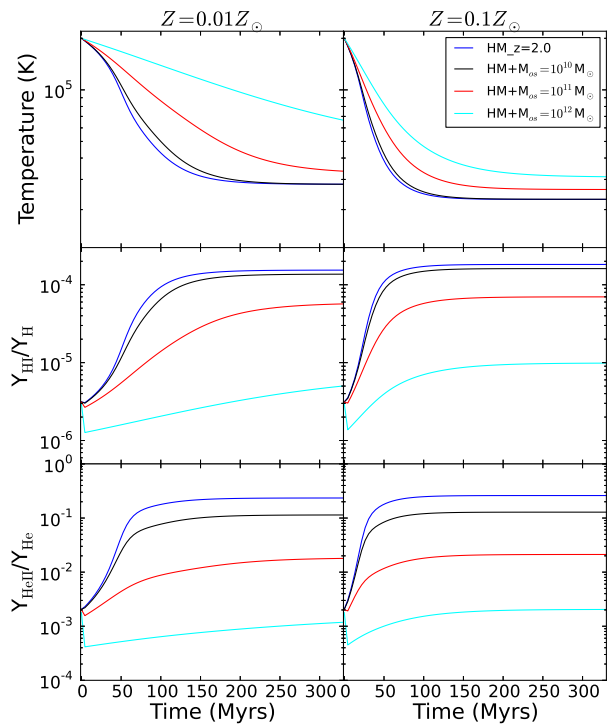}
\caption{ Same as Fig. \ref{fig:sfrx}, but in the radiation field created by
old ($>$200 Myr) stellar populations of 3 different masses.  The cooling 
time and the equilibrium temperature increases in the presence 
of a local radiation field from old stars, although not to such as extent as in radiation field from star formation.}
\label{fig:osr}
\end{center}
\end{figure}

Fig. \ref{fig:osr} shows the evolution of the same particles in the old star
radiation field.  Old stars have a smaller effect than young stars that include 
X-rays, especially in high 
metallicity gas.  The HeII fractions (bottom panels) 
are systematically lower in the old star radiation field (Fig. \ref{fig:osr}) 
than in the young star field (Fig. \ref{fig:sfrx}) because the flux at energies
just above the helium ionising edge is higher. 
While old stars ionise helium, they do not ionise metals,  
so the gas cools faster at high metallicities than it does in young star radiation field.

\section{Cosmological simulation using local photoionisation feedback}
\label{sec:simulation}

We want to study the effect of local ionising radiation on the whole
process of galaxy formation in a cosmological context. 
For this purpose, we implement local photoionisation
feedback in {\sc gasoline}.
The version of \textsc{gaoline} is the same described in \citet{2013MNRAS.428..129S}
with the changes to cooling that have been described in the previous section.

\subsection{Simulation physics}
We select a galaxy from the 
McMaster Unbiased Galaxy Simulations (MUGS, \citealt{2010MNRAS.408..812S}) 
that has a  halo mass of $6.4\times10^{11}$ M$_\odot$.  With the feedback 
prescription used in \citet{2013MNRAS.428..129S}, the star formation rate of 
the galaxy increases steadily to 12 M$_\odot$ yr$^{-1}$ at $z=0$.  
This galaxy was chosen to show the maximium effect of the local 
photoionisation field. The galaxy has a gas mass resolution of 
$2 \times 10^5$ M$_{\odot}$ with a gravitational force softening of $310$ pc.

The star formation and feedback model is the same as the fiducial simulation 
from the \citet{2013MNRAS.428..129S} parameter study.
The star formation efficiency, $c^\star$, is 0.1, and the density threshold is
9.3 cm$^{-3}$. 
The stellar population that forms follows a \citet{2003PASP..115..763C} initial stellar mass
function.  Stellar feedback comes from type II supernovae (SNII) and pre-supernova
energy (early stellar feedback, ESF hereafter).
Each SNII explosion ejects $E_{SN}=10^{51}$ erg of thermal energy into 
the surrounding ISM, which has its cooling delayed according to the blastwave
solution presented in \citet{2006MNRAS.373.1074S}. This cooling delay
typically lasts several Myr. Type II and Ia supernovae chemically enrich the ISM according
to a detailed stellar evolution model as described in \citet{2013MNRAS.428..129S}.
Energy injection before SNII explosions
(``early stellar feedback'') is modelled using
10\% of the bolometric luminosity from young stars, an amount comparable to
the UV flux.  This energy is deposited directly into the gas surrounding the young
stars as thermal energy, and then is rapidly radiated away.

We add our local photoionisation feedback to these stellar feedbacks.
As outlined in \S \ref{sec:cs}, the contribution from stars are summed 
based on their distance from the gas particle for which we want to compute
the incoming radiation.
Close sources (stars) are treated as individual sources, while distant
one are grouped based on the gravity tree.  A detailed description
of this method will be provided in Woods et al. (in prep).

We emphasize that the simulations presented here are preliminary.  
One concern is that we include the radiation energy from 
young stars twice, both as a photoionisation source and as a source
of thermal energy for the early stellar feedback.  Though the details of
the pre-supernova thermal energy input would change, the energy 
could also be the hot gas created by stellar winds.  In future models, we hope to
present a more consistent picture.  For now, we leave our model as similar
as possible to our previous simulations so that the effect of
including local photoionising radiation on the formation of 
the disc is apparent.  

A second concern is that the effect of low energy photons in our 
simulations may be overestimated because the outer regions of high 
density gas clouds will shield the inner regions from the radiation field. 
This `self-shielding' is important in gas with a density higher than 
$0.1$ cm$^{-3}$  \citep{2013arXiv1307.0943C}.
However, we have shown that the x-ray photons have the largest effect on 
cooling and these photons have a very small interaction cross section. 
So, ignoring self-shielding effects may be a reasonable first approximation.

\subsection{Simulation results}
\label{sec:simresults}

We perform two simulations of the galaxy, one including the HM UV background only 
(hereafter, HM), and one adding local photoionisation feedback (LPF) to 
the UV background (HM+LPF).   

\begin{figure}
\begin{center}
\includegraphics[scale=0.47]{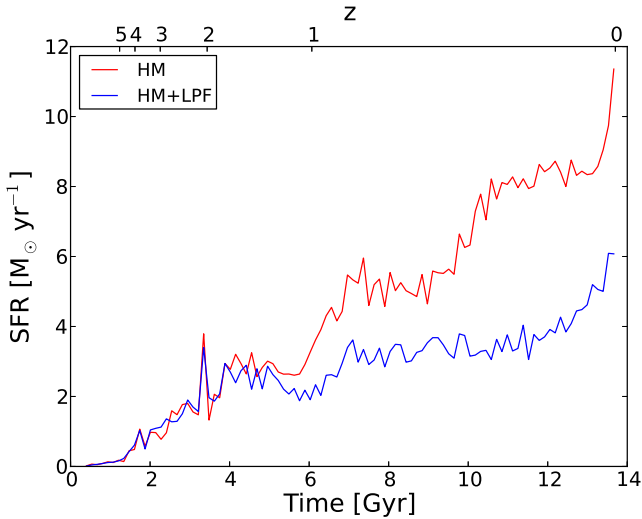}
\caption{A comparison of the star formation histories for the two simulations. 
The star formation rates are similar until $z\sim1.5$ and then they begin to diverge.}
\label{fig:sfh}
\end{center}
\end{figure}

Fig. \ref{fig:sfh} shows the star formation history of the two simulations.   
The star formation histories are nearly identical until $z\sim1.5$, 
after which the they diverge.  The star formation rate in the HM run 
(red curve) steadily increases to $12$ M$_{\odot}$ yr$^{-1}$ at $z=0$. The 
HM+LPF simulation (blue curve) maintains a steady star formation
rate of $\sim\,4$ M$_{\odot}$ yr$^{-1}$ from $z=1.5$ until $z=0$. 
The reduced star formation results in $\sim40\%$ less stellar mass than in
HM at $z=0$. 

\begin{figure*}
 
 \centering
 \subfigure[HM ]{ 

  \includegraphics[scale=0.42]{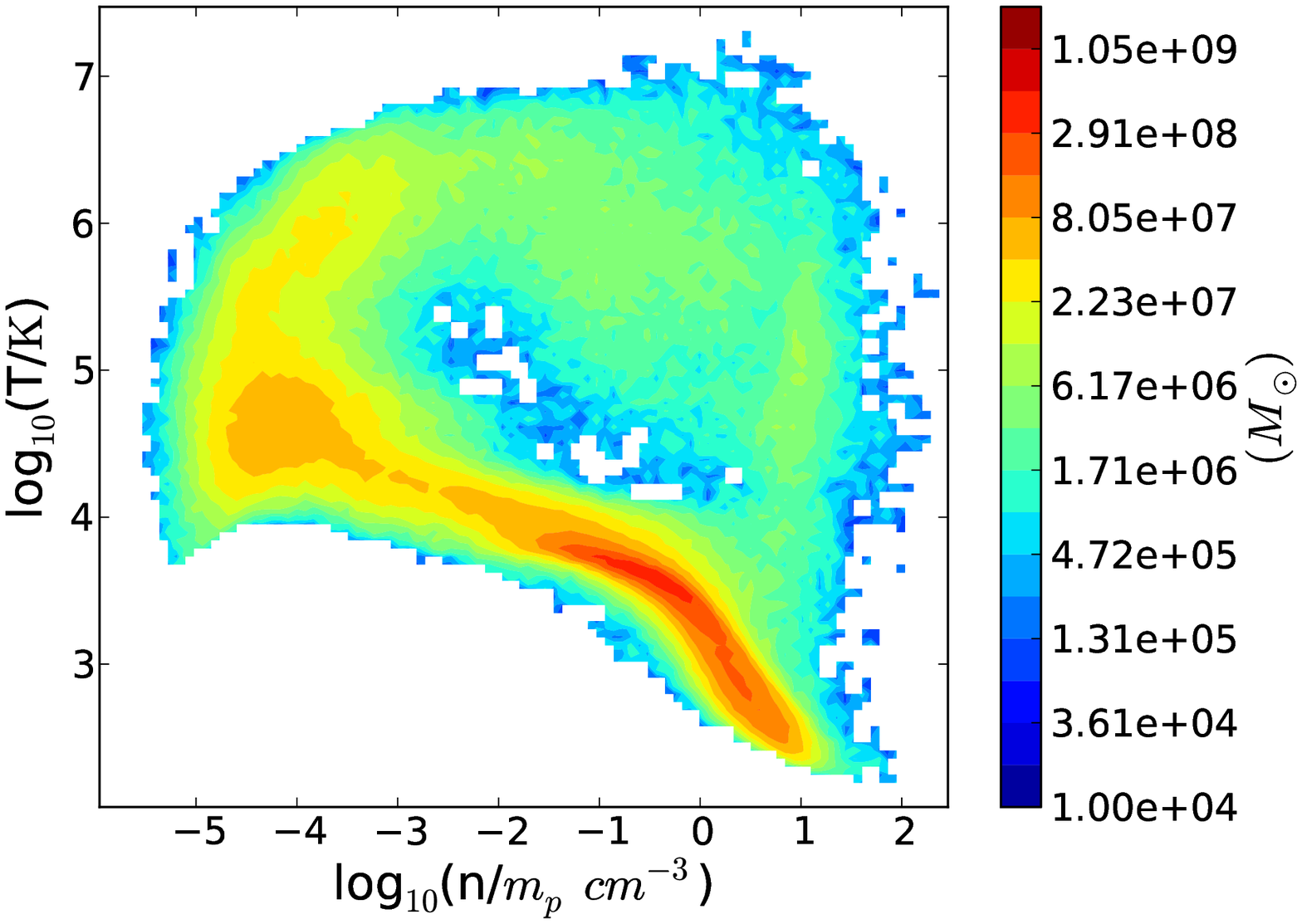}
   }
 \subfigure[HM+LPF]{
  \includegraphics[scale=0.42]{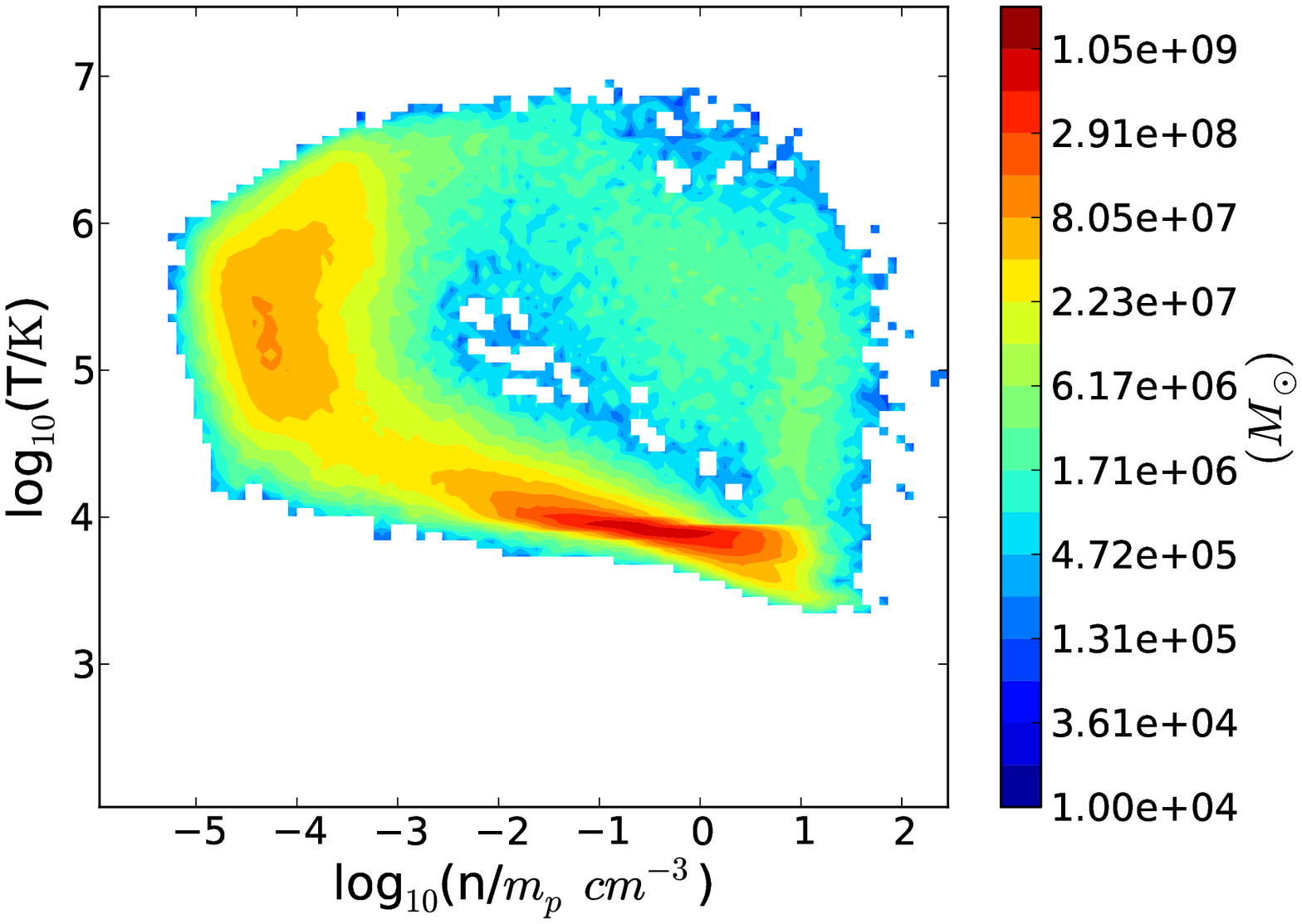}
   }
 \caption{Temperature--density phase space diagrams present in 
 a phase bin, 0.1 dex in temperature by 0.1 dex in density at $z=0$.  Three effects of local 
 photoionization are visible: 
 (i) High density, low temperature gas is absent in the presence of the local radiation field, 
(ii) The mass of gas accreting from the halo to the disc is reduced (gas channel from top left to bottom right) and 
(iii) the temperature of the hot halo gas around the galaxy is significantly higher.}
\label{fig:phase}
\end{figure*}

To understand the physics behind the lower star formation rate in HM+LPF, 
Fig. \ref{fig:phase} shows
a comparison of the distribution of the gas in temperature--density phase 
space of the two simulations at $z=0$.  Three differences are apparent 
between the simulations: the mean temperature of
low density halo gas, the amount of gas cooling out of the halo onto 
the disc, and the absence of very low temperature gas in the disc of the 
galaxy.  

The difference between the hot, diffuse halo gas in the region bounded by  $10^5<$T(K)$<10^6$  and $log(n/m_p cm^{-3}) <$ $-3.5$ 
is striking.
With local photoionising feedback (LPF), most of the low density gas mass is between $10^5$ and $10^6$ K.  
Without LPF, a large fraction of the halo gas has cooled to
a phase that is intermediate between the cool, dense disc and the 
hot, diffuse halo and the gas temperature lies between $10^4$ and $10^5$ K. 
Even a small amount of ionising radiation from the local sources has a 
big effect on the gas cooling rate, so that gas stays hot longer.
The high temperature of the hot halo gas provides pressure support 
against the galaxy's gravitational potential and hence reduces the 
gas accretion rate onto the disc.  Thus, there is less gas in the region 
between the hot, low dense halo gas and the cold dense disc in HM+LPF
than in HM.  

Another factor that causes more cooling in HM is the positive feedback
that the metal enrichment from the higher star formation rate in HM causes.  
HM starts with a marginally higher accretion rate than HM+LPF, which ejects
more metals into the hot halo, which cause the gas to cool faster further enhancing
the accretion rate, and consequently makes more stars.  
The higher halo gas temperature indicates the global nature of 
LPF and underlines the importance of propagating the radiation field from 
the local sources throughout the entire volume of the simulation box. 

\begin{figure}
\begin{center}
\includegraphics[scale=0.47]{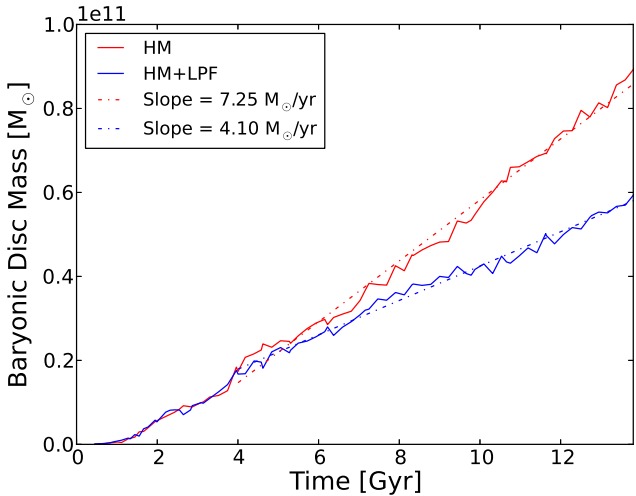}
\caption{The evolution of the baryonic disc mass of the galaxy as a function of time, for the simulations with (blue curve) and without (red curve) local photoionization.}
\label{fig:gasacc}
\end{center}
\end{figure} 

Fig. \ref{fig:gasacc} shows the reduction in gas accretion rate onto the disc more explicitly.
The baryonic disc mass evolves similarly in the two simulations until $\sim4$ Gyr when they diverge as HM adds mass at a faster rate, leaving HM+LPF lighter than HM. 
The slope of the baryonic disc mass evolution gives a 
rough gas accretion rate onto the disc (modulo outflows and stellar accretion).  The dot-dash line represents a best linear fit of the gas accretion rate 
after $\sim4$ Gyr.  For the HM simulation the slope of this line is 
$7.25~ \rm{M}_{\odot} \rm{yr}^{-1}$, while the HM+LPF simulation has a slope
of $4.10~  \rm{M}_{\odot} \rm{yr}^{-1}$. 

The increased cooling rate and equilibrium temperature of the halo gas due to LPF affects star formation by 
reducing the gas accretion onto the disc. This is different from feedback 
mechanisms local to star forming events that rely on blowing gas out of the 
disc.  LPF is rather a {\it preventive} feedback mechanism which reduces the 
need for artificially high levels of feedback that can destroy the galaxy 
disc (\citealt{2011MNRAS.410.1391A}; \citealt{2013arXiv1308.6321R}).  LPF also provides a natural 
and non-violent mechanism to keep the disc lighter and prevent disc 
instabilities from driving gas to the center.  

\begin{figure*}
\begin{center}
\includegraphics[scale=1.0]{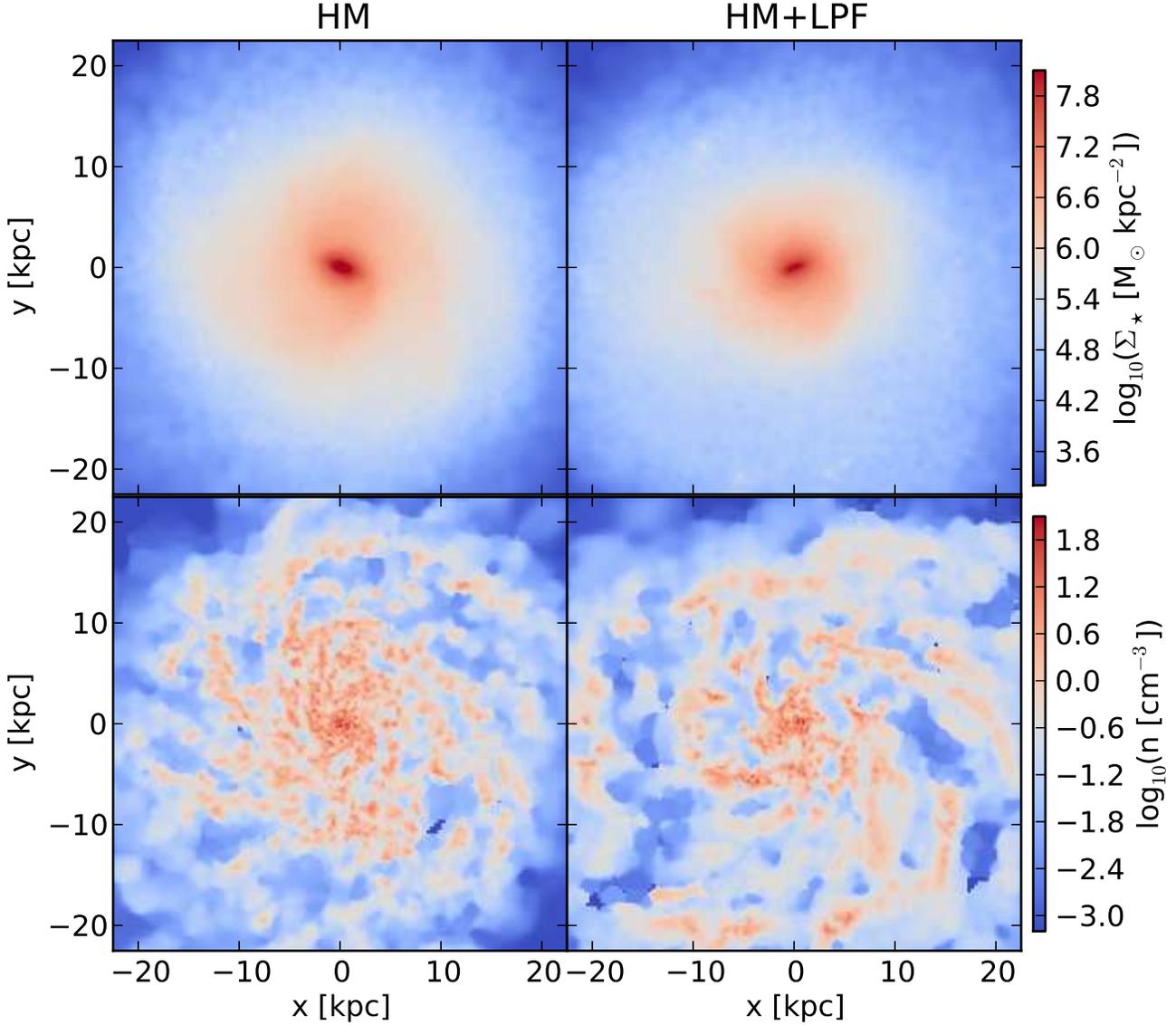}
\caption{The face on projection view of the stellar (top panels) and gaseous (bottom panels) of the MW like galaxy simulations with the HM (left panels) and HM+LPF (right panels) models.}
\label{fig:fcgas}
\end{center}
\end{figure*}

 Fig. \ref{fig:fcgas} shows face-on images of the stellar and gaseous components of the simulated galaxies. The top panels show the stellar distribution and the bottom panels the gas distribution for the HM (left panels) and HM+LPF (right panels) simulations. The HM simulation shows a large stellar bulge and a high concentration of gas in the center. The HM+LPF simulation has a smaller stellar bulge with generally less mass in stars in the disc. The gas distribution in the HM+LPF simulation is more extended and less concentrated than in the HM simulation.  

\begin{figure}
\begin{center}
\includegraphics[scale=0.47]{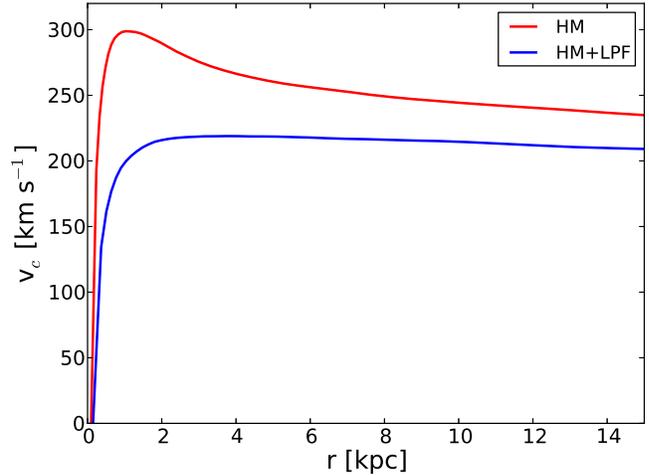}
\caption{The rotation curves of the galaxy in the HM (red) and HM+LPF 
(blue) runs at $z=0$.}
\label{fig:rc}
\end{center}
\end{figure}

The central concentration of stars and gas in the HM simulation is reflected in the $300$ \kms central peak of the galaxy rotation curve shown in
Fig. \ref{fig:rc}.  The HM+LPF simulation creates a slowly rising rotation curve that has
a rotation velocity of $200$ \kms from near the center to the edge of the disc, clearly showing  that it is much lighter than the HM disc. 

The difference in stellar and gaseous distributions and subsequently in the rotation curves can be explained in part due to the reduced gas accretion onto the disc, but also with the reduction in amount of
cold gas in the HM+LPF. The HM simulation has a tail of gas at low 
temperatures ($<1000$ K) and high densities ($n>1$ cm$^{-3}$).  This tail 
is absent from the HM+LPF run since the additional radiation 
fields raise the equilibrium temperature of the dense gas. Thus, 
the average temperature of the disc gas is higher and has higher pressure that
stops the disc from fragmenting and forming stars.  However, we reiterate that the effect of low energy photons might be overestimated, because we do not impose any criterion for self-shielding in high density gas present in the disc.

\section{Conclusions}
\label{sec:conclusion}

We tested a novel method for including the effects of local photoionising
radiation fields in galaxy formation simulations.  Previously, the expense
of including detailed radiative transfer meant that simulations that
tried to include the effect of photoionisation could only be evolved until
$z\sim4$.  
While such simulations are extremely useful to study the early evolution of the
Universe and the reionisation epoch they are not able to address the effect of
a local radiation field on galaxy evolution.
The local radiation field should have a significant importance in regulating how much
 gas cools onto discs and fuels star formation.  Currently, in many simulations, the only radiation field considered is the UV background
against which the galaxy might otherwise be self-shielded  
(\citealt{2007MNRAS.381L..99P}; \citealt{2009ApJ...703.1416F}).  The local field might be very
different from the uniform background and must be included
in realistic cosmological simulations of galaxy formation.

Our method uses the existing optimizations that quickly solve gravity
to include the $r^{-2}$ attenuation effect,
a more detailed description of our method will be presented in 
Woods et al. (in prep).
Our current treatment uses the optically thin limit, which
provides an upper limit on the effect of radiation. The optically thin 
approximation is valid for the x-ray photons that we include in our model
of the young star radiation field.  X-ray photons are not strongly
absorbed in typical ISM conditions. 
However, we will relax the optical thin approximation in future work.

Our radiation field includes the black body emission from young stars, 
along with the x-rays that massive stars produce in their winds and supernova
explosions, as well as UV flux from post-AGB stars in old stellar 
populations.  Our simple treatment of absorption assumes that 
5\% of the flux around the Lyman limit escapes from
clusters of young stars, while old stars have an escape fraction of unity since they 
likely moved out of their birth cocoons into an optically thinner environment. 
Otherwise, the radiation field is only attenuated as an inverse square of distance. 
These physically motivated assumptions allow us to study the effect of a local ionising 
radiation field on cooling rates of gas within the galaxy, throughout cosmic time.

As a test of the cooling and to develop physical intuition about 
how photoionisation affects the cooling rate, we provide 
simple examples of the evolution of a single gas parcel embedded in the 
radiation field.  
We show that a radiation field from young stars that includes soft x-rays
can increase the equilibrium temperature of the gas significantly and 
prolong cooling times \citep{2010MNRAS.403L..16C}.

We then use our local photoionisation scheme in the SPH code {\sc gasoline} to investigate the effect of local ionizing 
radiation fields in full cosmological simulations of a Milky Way-like galaxy.
We simulate the galaxy with and without the local radiation field.  
We find that the radiation field reduces star formation after $z\sim1.5$, 
and results in $\sim$ 40\% less stellar mass. 
The reduced star formation is due to a combination of factors. 
The hot, diffuse halo gas surrounding the disc has a higher temperature when the 
local photionising field is considered because a small amount of ionising 
radiation from local sources has a big effect of the gas cooling and heating rates 
at low densities, which in turn raises the equilibrium temperature of the gas.
This increased temperature of the hot halo gas provides pressure support to the halo gas against the gravitational potential of the galaxy 
and hence reduces the gas accretion rate onto the disc. This coupling of the local radiation 
field to the gas cooling in the host galaxy provides a {\it preventive} 
feedback mechanism that reduces the gas accretion to the central regions of 
the galaxy, regulating star formation. 

The local ionising radiation field also eliminates high density-low 
temperature gas by raising the equilibrium temperature 
of dense gas in the disc.  The higher average temperature of the disc gas 
provides pressure support to the gaseous disc that stops the disc 
from fragmenting and forming stars. 
All these effects on the gas distribution by the local radiation field causes the HM+LPF run to form a light and more stable stellar disc, 
which has a slowly rising rotation curve which peaks at $200$ \kms, 
consistent with observations of Milky Way-like galaxies.

We plan to extend this initial study to a broader galaxy mass range 
(from dwarfs to massive ellipticals) and to improve the parametrization
of our radiative transfer scheme in forthcoming work(s). While this result is still preliminary and based on a single simulation, it shows
the importance of self-consistently including local photoionisation feedback 
in simulations aimed at reproducing realistic galaxies.

\section*{Acknowledgements}

The analysis was performed using the pynbody package
(\texttt{http://code.google.com/p/pynbody}), which was written by Andrew Pontzen in addition to the authors.
RK, GS and AVM acknowledge financial support through the  Sonderforschungsbereich SFB 881 ``The Milky Way System''
(subproject A1) of the German Research Foundation (DFG). 
Numerical simulations were performed on the Milky Way supercomputer,
funded by the Deutsche Forschungsgemeinschaft (DFG) through Collaborative Research Center (SFB 881) 
"The Milky Way System"
(subproject Z2), hosted and co-funded by the J\"ulich Supercomputing
Center (JSC) along with \textsc{theo} cluster of  the
Max-Planck-Institut f\"ur Astronomie at the Rechenzentrum in Garching;
and the clusters hosted on \textsc{sharcnet}, part of ComputeCanada.  We greatly appreciate
the contributions of these computing allocations. JFH acknowledges generous support from the Alexander von Humboldt
foundation in the context of the Sofja Kovalevskaja Award. The
Humboldt foundation is funded by the German Federal Ministry for
Education and Research.

\bibliographystyle{mn2e}
\bibliography{radcool.bib}

\clearpage

\end{document}